\documentclass[10pt,conference]{IEEEtran}
\IEEEoverridecommandlockouts
\usepackage{cite}
\usepackage{amsmath,amssymb,amsfonts}
\usepackage{algorithmic}
\usepackage[caption=false,font=footnotesize]{subfig}
\usepackage{graphicx}
\usepackage{textcomp}
\usepackage{xcolor}
\usepackage{siunitx} 
\usepackage{url} 
\usepackage{color,soul}
\usepackage{todonotes}
\usepackage{enumitem}
\usepackage{subcaption}
\setlist{nosep}

\begin{document}

\newcommand{\pavel}[1]{\todo[color=red!20]{Pavel: #1}}
\newcommand{\kyriakos}[1]{\todo[color=blue!20]{Kyriakos: #1}}
\newcommand{\liliana}[1]{\todo[color=green!20]{Liliana: #1}}

\title{
ACTING: A Platform for Cyber Ranges Federation 
\thanks{
This work was supported by the ACTING Project, Co-funded by the European Union through the European Defence Fund (EDF) and in part by the European Union’s Horizon 2020 research and innovation programme under Grant Agreement No 739551 (KIOS CoE - TEAMING) and from the Republic of Cyprus through the Deputy Ministry of Research, Innovation and Digital Policy. Views and opinions expressed are however those of the author(s) only and do not necessarily reflect those of the European Union or European Commission. Neither the European Union nor the granting authority can be held responsible for them.
}
}

\author{\IEEEauthorblockN{
Kyriakos~Christou\IEEEauthorrefmark{1}, Maria~Michalopoulou\IEEEauthorrefmark{1}, 
Stefano~Taggi\IEEEauthorrefmark{2}, 
Matteo~Merialdo\IEEEauthorrefmark{2}, 
Nikolai~Stoianov\IEEEauthorrefmark{3},\\
Vasilis Ieropoulos\IEEEauthorrefmark{1},
Theofanis~Eleftheriadis\IEEEauthorrefmark{1},
Philippos~Isaia\IEEEauthorrefmark{1}, 
Eleni~Darra\IEEEauthorrefmark{5}, 
Ilias~Koritsas\IEEEauthorrefmark{5},\\
Antonis~Voulgaridis\IEEEauthorrefmark{5},
Giorgos~Rizos\IEEEauthorrefmark{5}, 
Dimitris~Kavallieros\IEEEauthorrefmark{5},
Stefanos~Vrochidis\IEEEauthorrefmark{5},
Konstantinos~Votis\IEEEauthorrefmark{5},\\
Liliana~Medina\IEEEauthorrefmark{6},
João~Camacho\IEEEauthorrefmark{6},
Tim~Gerling\IEEEauthorrefmark{7},
Aimilia-Bantouna\IEEEauthorrefmark{10},
Pavel~Varbanov\IEEEauthorrefmark{8},
George~Sharkov\IEEEauthorrefmark{8},\\
Christos~Laoudias\IEEEauthorrefmark{1},
Jos\'e~Borges\IEEEauthorrefmark{9},
Maria~K.~Michael\IEEEauthorrefmark{1}\IEEEauthorrefmark{4}
}

\IEEEauthorblockA{
\IEEEauthorrefmark{1}KIOS Research and Innovation Center of Excellence, University of Cyprus, Cyprus;
\IEEEauthorrefmark{2}Nexova, Belgium;\\
\IEEEauthorrefmark{3}Bulgarian Defence Institute, Bulgaria;
\IEEEauthorrefmark{4}Department of Electrical and Computer Engineering, University of Cyprus, Cyprus;\\
\IEEEauthorrefmark{5}Information Technologies Institute, Centre for Research and Technology Hellas, Greece; 
\IEEEauthorrefmark{6}VisionSpace Portugal, Portugal; \\
\IEEEauthorrefmark{7}VisionSpace Technologies GmbH, Germany;
\IEEEauthorrefmark{8}European Software Institute, Cybersecurity Laboratory, Bulgaria;\\
\IEEEauthorrefmark{9}CINAMIL -- Centro de Investigação, Desenvolvimento e Inovação da Academia Militar, Lisboa, Portugal;\\
\IEEEauthorrefmark{10}WINGS ICT SOLUTIONS SA, Athens, Greece
}

}

\newtheorem{definition}{Definition}[section] 

\IEEEoverridecommandlockouts
\makeatletter
\def\@IEEEaftertitletext{\vspace{-0.7cm}}
\makeatother

\maketitle

\begin{abstract}
Cyber Defence (CD) training requires interoperable cyber-range environments capable of supporting complex, multi-domain exercises across distributed infrastructures. This paper presents three main contributions addressing this challenge. First, we introduce the Exercise Description Language – First Generation (EDL-FG), a structured language for formally describing cyber-range training services and exercises. EDL-FG captures both the technical infrastructure required to emulate ICT/OT environments and the scenario logic governing cyber events, injects, and participant interactions, enabling interoperable and automated scenario deployment across federated Cyber Ranges (CRs). Second, the ACTING platform introduces automated PE and scoring mechanisms that assess trainee actions during exercises through coordinated data collection and analysis across participating CRs. Third, the platform enables multi-domain cyber training scenarios that combine civilian and military operational contexts. Building upon federation capabilities established under the H2020 ECHO project, ACTING demonstrates how interoperable scenario description and automated evaluation support scalable and realistic CD training.
\end{abstract}

\setlength{\textfloatsep}{-0pt}

\section{Introduction}
\label{sec:intro}
CRs are established technologies for the controlled and realistic emulation of cyber incidents and defensive operations, supporting education, testing, and training in research and operational practice~\cite{ECSO2020, Chouliaras2021}. Conventional CRs typically target a single domain or cybersecurity aspect, whereas contemporary cyber incidents span interconnected domains and infrastructures, where dependencies and cascading effects shape attack dynamics and response strategies. However, developing and maintaining multi-domain CR environments requires substantial expertise and resources. Federated CRs address this limitation by interconnecting independently developed CRs under common operational and technical agreements, enabling multi-domain scenarios within a unified environment~\cite{Mudassar2025}.

Recent research and CD initiatives highlight both the operational relevance and architectural complexity of Federated CRs. Federation is a multi-dimensional challenge involving technological, operational, and organizational aspects~\cite{Lal2026}. Beyond technical interconnection~\cite{Virag2021}, key challenges include interoperability across heterogeneous CR infrastructures, scalability, resource management, and secure data exchange, including identity management and access control~\cite{Nicodeme2020}. Governance, trust, legal constraints, and sustainability also affect viable federated ecosystems, while prior work emphasizes common scenario description frameworks for technology-agnostic design and cross-platform execution~\cite{Park2022, Oikonomou2021}.

Although prior work demonstrates the feasibility of interconnectivity, interoperability, scalability, and runtime aspects of federated environments, it focuses on specific, separate solutions, such as scenario modelling and execution, infrastructure and resource details, automation, and script parsing~\cite{Yamin2022,Rizos2025}. However, complex multi-domain exercises require tighter integration of federation mechanisms with Situational Awareness (SA), PE, and realistic simulation capabilities. Challenges in scenario portability and harmonized operation across heterogeneous CR infrastructures further motivate more cohesive architectural frameworks~\cite{Katsantonis2023}. This work introduces the capability to jointly represent the CR topology, the scenario, and the associated scoring mechanisms described above.

The main contributions of this work are threefold:

\begin{itemize}[itemsep=0pt, topsep=2pt, parsep=0pt, partopsep=0pt]

\item \textit{EDL-FG}: A structured language to formally describe CR training services and exercises, specifying both the technical infrastructure required for simulation environments and the scenario logic governing cyber events and participant interactions. EDL-FG provides a common and interoperable representation that supports automated deployment, orchestration, and collaboration across federated cyber-range environments.

\item \textit{PE and Scoring through the ACTING Platform (AP)}: The AP introduces automated mechanisms for evaluating trainees during exercises. Through coordinated data collection and analysis across participating CRs, it enables performance assessment, scoring, and structured feedback for trainees and instructors, supporting improved understanding of participant behaviour and training outcomes.

\item \textit{Cross-Domain CR Training Scenarios}: The AP supports the development and execution of CR scenarios that combine civilian and military domains, enabling realistic multi-domain training exercises and cross-sector collaboration in federated environments.

\end{itemize}

The remainder of this paper is structured as follows. Section~\ref{sec:arc} outlines the architecture and key blocks of the AP. Section~\ref{sec:func} describes how these components operate together to realize the envisioned cybersecurity training environment, providing advanced features and functionalities. Section~\ref{sec:scen} demonstrates how the integrated tools and data available in the AP implement scenarios addressing various needs and gaps in the defence sector. Finally, Section~\ref{sec:conc} concludes the paper and outlines directions for future work.

\section{System Architecture}
\label{sec:arc}

\subsection{Requirements for Advanced Cyber Training}
\label{ssec:reqs}

The requirements for advanced cyber training environments can be summarised as follows:

\begin{itemize}[itemsep=0pt, topsep=2pt, parsep=0pt, partopsep=0pt]

\item Realistic and Adversarial Training Environments.  
CRs should replicate realistic threat conditions and enable controlled exposure to adversarial tactics against complex ICT and cyber-physical systems~\cite{Glas2023, Shin2024}.

\item Support for Interdependent Infrastructure.  
Training environments must simulate multidimensional attacks across interconnected systems and critical infrastructures.

\item Time-Critical Decision-Making Support.  
CD training should support rapid decision-making under operational pressure, reflecting the time-sensitive nature of cyber incidents~\cite{Glas2023}.

\item Federated and Interoperable Infrastructure.  
CRs should aggregate heterogeneous resources and ensure scenario portability across platforms to strengthen preparedness and resilience.

\item Support for Joint and Coalition Operations.  
Training platforms must support interoperability across organisations and domains to reflect real-world operational constraints.

\item Automated Evaluation and Continuous Learning.  
Advanced environments should include automated PE, user behaviour modelling, and structured feedback to improve cyber readiness.

\end{itemize}

\subsection{Overview of the ACTING Platform}
\label{ssec:overview}
\begin{figure}[htbp]
\centering
\includegraphics[width=\linewidth]{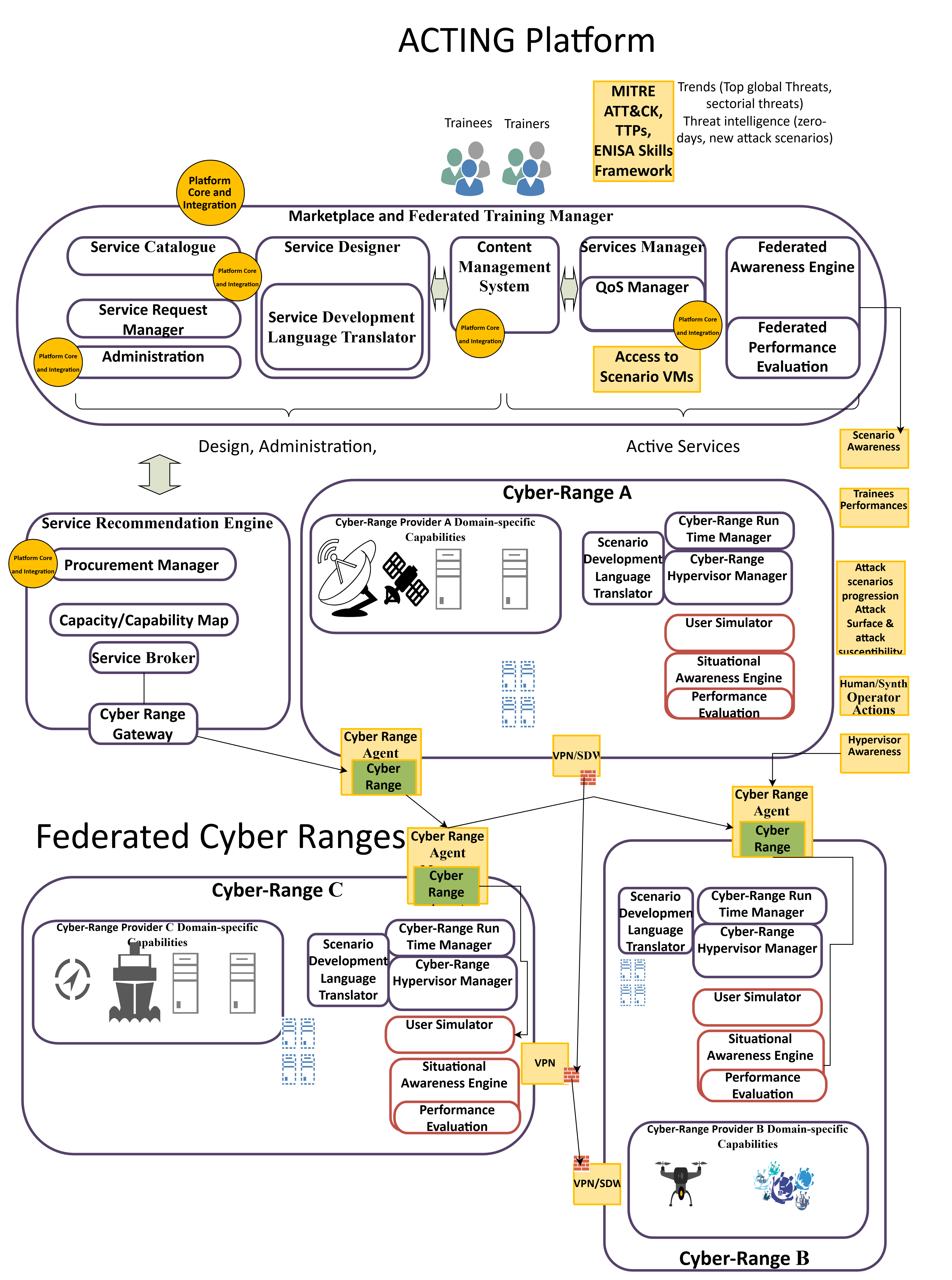}
\caption{ACTING High-level Design. }
\label{fig:actingHLD_2}
\end{figure}

The AP is an integrated digital environment that supports the discovery, customization, procurement, and consumption of cybersecurity services. It connects service providers and customer organizations through a unified framework, enabling access to Cyber Range (CR) scenarios, tabletop exercises, and training modules. As an orchestration layer, it allows providers to publish services and customers to discover, negotiate, and request them through structured workflows, while also supporting customized services delivered collaboratively by multiple providers. The high-level architecture of the AP is shown in Fig.~\ref{fig:actingHLD_2} and comprises the following building blocks.

Importantly, ACTING is neither a CR infrastructure nor a Learning Management System (LMS). Rather, it acts as an aggregation and coordination layer connecting independent providers and resources. Its value depends on service providers actively publishing offerings, positioning ACTING as an enabling marketplace and orchestration environment for distributed cybersecurity training and exercise delivery.

\subsection{Main Building Blocks}
\label{ssec:blocks}
The ACTING Consortium is developing the ACTING Training and Simulation Platform as an integrated environment for advanced cyber capability development and exercise orchestration. It builds on core research and technical themes that enhance the effectiveness and scalability of cyber training, including CR SA for monitoring complex exercises and automated PE for objective, data-driven assessment of participant actions and outcomes. The platform also incorporates user behaviour simulation to model human and organisational interactions within CR scenarios. Another key contribution is the Scenario Development Language, which standardises and accelerates scenario creation and portability. In addition, ACTING adopts a federated approach to enable interoperable, multi-sector CR exercises. Together, these capabilities form the technological foundation of the AP.

\subsubsection{Building Block 1} Service Catalogue (SC) and Service Request Manager (SRM) 
\label{sssec:buildingBlock1}

The SC is the central repository of platform services and the main entry point for end users. It allows Service Providers to publish and manage services, while Service Requesters can browse, search, and initiate requests. The SC supports federated publishing across providers, advanced filtering, search, and metadata tagging based on frameworks such as NICE and SFIA. It integrates with the Administration Component, Service Designer, QoS Manager, Procurement Manager, and Notification Manager, and forwards selected services to the SRM.

The SRM manages the lifecycle of service requests and acts as the negotiation and approval hub between Service Requesters and Service Providers. It supports workflows from request initiation to approval, communication, and contract finalization through integration with the Procurement Manager. The SRM also interacts with the Service Broker, Notification Manager, and Services Manager, providing request status visibility, secure session handling, and role-based access control for the transition from service discovery to execution.

\subsubsection{Building Block 2} Service Designer (SD)
\label{sssec:buildingBlock2}

The SD is a core platform component that enables users to define and configure new services or service requests according to their role. Service Providers use it to create new offerings, while Service Requesters use it to initiate requests or customize catalogue entries by adjusting size, difficulty, duration, and features. A guided wizard supports users from high-level descriptions to technical and organizational details, enabling both simple and complex service definitions. At its core, the SD uses the concept of a Unit, which represents a basic service building block. A service may include one or more Units, supporting both simple and composite services. The following unit types are currently supported:

\begin{itemize}

\item Content-based Training: A structured learning module or training program for developing cybersecurity skills, knowledge, or competencies. It may include instructional content, learning materials, and assessments, and can be used as standalone training or preparation for advanced exercises.

\item CR Scenario: An immersive, hands-on simulation of cybersecurity incidents, such as malware infection, unauthorized access, or system failure. Participants interact with a virtualized environment to detect, respond to, and mitigate threats in real time.

\item Tabletop Exercise: A discussion-based exercise that simulates cybersecurity events through structured dialogue. Participants work through incident response, communication, and decision-making processes to improve coordination and strategic awareness.

\end{itemize}

By combining Units, the SD supports customized, modular services tailored to different training and operational objectives, improving flexibility, reusability, and alignment between providers and requesters. For CR Scenario units, users work with a visual canvas that serves as both a design workspace and live configuration tool. The drag-and-drop interface provides building blocks such as virtual machines, network segments, traffic generators, scoring modules, and access policies, which can be arranged to create realistic exercises. Each change updates a live model that functions as both a human-readable blueprint and a machine-ready deployment manifest. The canvas also abstracts multi-site deployments, allowing users to design across federated CRs in a single unified space while the platform manages orchestration, resource allocation, locality constraints, validation, and scheduling. Overall, the SD enables providers and requesters to create and adapt CR scenarios visually and efficiently without writing infrastructure code.

\subsubsection{Building Block 3} Service Development Language Translator (SDL-T)
\label{sssec:buildingBlock3}

The SDL-T component transforms high-level service specifications into executable CR scenarios using the EDL-FG framework, as described in Section IV-C. This semantic transformation maps abstract scenario definitions into deployable operational configurations. Through EDL-FG validation and the ACTING workflow, it supports iterative refinement and verification of scenario elements. It also improves collaboration among technical developers, cybersecurity experts, domain specialists, and training designers through a common formalised representation that reduces ambiguity. Service components are formally described, including IT/OT infrastructures, virtual and physical assets, human and simulated actors, data flows, dependencies, roles, behaviours, interactions, and constraints for cyber-event orchestration and participant engagement. These descriptions are translated into machine-readable artefacts for orchestration and evaluation, ensuring consistency between design and execution and enabling reproducible, scalable, and measurable cyber-range scenarios.

\subsubsection{Building Block 4} Service Recommendation Engine (SRE)

The SRE forms the backbone of the AP during the pre-execution phase. It enables CR providers to register and maintain structured information on their infrastructure capabilities, resource capacities, and availability. Using this information, the SRE manages the service-request lifecycle, including analysis, resource allocation, and activation. It also supports matchmaking by identifying suitable alternatives when requested resources are unavailable, while ensuring alignment with requester requirements and constraints. The SRE consists of two main components: the \textit{Capacity-Capability Map (CCM)} and the \textit{Service Broker (SB)}.

\begin{itemize}

\item Capacity-Capability Map (CCM)
The CCM stores and manages CR data, including available services, capabilities, capacity, and reservation status. It enables users or requesters to design a desired CR service configuration and verifies whether it can be offered based on available capabilities, capacity, and reservations.
\begin{definition}[Cyber Range Capability]
A CR Capability is a \textbf{qualitative} descriptor of the operational scope, purpose, and functional properties of a CR.
\end{definition}
\begin{definition}[Cyber Range Capacity]
A CR Capacity is a \textbf{quantitative} descriptor of measurable static CR resources.
\end{definition}

\item Service Broker (SB)
The SB processes service requests and manages their lifecycle within the platform. It also acts as a recommendation engine by analysing requests and checking whether the required resources are available. After receiving a request from the Service Request Manager, the SB queries the CCM to identify the most suitable CR or combination of CRs for federated execution. If validation succeeds and the invoice is generated, the resources are reserved in the CCM and the request is forwarded to the CR provider(s) for activation. If validation fails due to insufficient resources or unavailability, the SB proposes ranked alternatives based on predefined evaluation metrics.
     
\end{itemize}

\subsubsection{Building Block 5} Federated Awareness Engine (FAE)
\label{sssec:buildingBlock5}

The FAE provides SA capabilities within the AP by collecting and analysing data generated during cyber-range exercises. The component correlates scenario events, participant actions, and contextual information to support the assessment of team performance and decision-making throughout the exercise lifecycle. By employing a set of time-aligned metrics, the engine enables the evaluation of SA levels across federated training environments.

\subsubsection{Building Block 6} Federated Performance Evaluation (FPE)
\label{sssec:buildingBlock6}

The FPE framework of the AP enables automated assessment of trainee performance in cyber-range training exercises, including scenarios executed across federated CRs. It supports the collection and analysis of participant actions during exercises, providing performance metrics and structured feedback to assist instructors in evaluating trainee problem-solving and decision-making capabilities. The framework consists of two main components: the PE module deployed at each participating CR, which monitors trainee activity locally, and the FPE module, which aggregates and analyses data across multiple ranges. Through this architecture, ACTING provides a unified view of trainee performance, enabling automated reporting, visualisation of results, and recommendations for further training.

\subsubsection{Building Block 7} Learning Management System 
\label{sssec:buildingBlock7}

The LMS supports the management and execution of training activities within the AP. It enables trainers to configure scenarios, define participant roles, and manage simulated user behaviour in cyber-range exercises. In collaboration with the US, it supports simulated user profiles, such as Blue, Red, and Grey users, schedules their actions, and orchestrates their activities during execution. The LMS also enables monitoring and interaction with ongoing exercises through integration with other ACTING components. Further details on its architecture and functionality are presented in a later section.

\subsubsection{Building Block 8} Services Manager (SM)
\label{sssec:buildingBlock8}

The SM is the central orchestration component of the ACTING cybersecurity training platform, responsible for managing the lifecycle of Negotiated Services across the Pre-Execution, Execution, and Post-Execution phases. Its core functions include access control, service configuration, monitoring, and execution coordination. The SM integrates with key platform subsystems—including the CR infrastructure, LMS, SA, FPE, CR Agent Manager (CRAM), Scenario Description Language (SDL), Services Catalogue, and the Service Request Manager—through well-defined interfaces and workflows. Configuration of Simulated Users, PE metrics, and SA parameters is facilitated through the SM in collaboration with the respective components. Additionally, the SM supports user and team management, action timeline configuration, virtual machine integration via Guacamole, and scenario topology visualisation, enabling near real-time synchronization and role-based access control across federated training environments.

\section{Main Features and Functionalities}
\label{sec:func}

This section presents the main features and functionalities of the AP that support advanced cybersecurity training in federated cyber-range environments. In particular, it describes the SA feature for monitoring and understanding cyber-range activities, the PE feature for automated assessment of trainee performance, and the Service Description Language Translator together with the EDL-FG, which enable the definition and interoperability of cyber-range scenarios across different infrastructures.

\subsection{Situational Awareness (SA)}
\label{ssec:datavisualzation}
The SA component is a core element of the CR and AP, collecting and processing data to analyse the situational understanding of Service Consumers, or trainees, and their teams during service execution~\cite{Alavizadeh2022}. It supports informed decision-making by recording and correlating scenario events, participant actions, issued hints, and contextual information along the scenario timeline. Using predefined metrics based on team composition, participant roles, and scenario objectives, the SA component evaluates SA levels and identifies potential team strengths and weaknesses.

The SA assessment also considers information provided by the CRs before exercise execution, including action-prioritisation guidelines and best practices expected from trainees and teams. The analysis incorporates performed and reported actions, achieved goals, followed strategies, and used hints, enabling a comprehensive assessment of participant performance and SA behaviour.

The Federated Analytics Engine (FAE) aggregates and analyses data metrics from all federated CRs and scenarios, including single-range and federated exercises. The FA component operates within the central AP and interacts with components such as the SDL-T and Service Manager. Architecturally, FA includes two subcomponents: the FAE, which performs backend data analysis, and the FA Visualisation module, which provides an interface for accessing insights and interacting with the system.

\subsection{Performance Evaluation (PE)}
\label{ssec:dashboard}
Evaluating trainee performance in federated CR scenarios is challenging due to the volume and diversity of generated data~\cite{vsvabensky2022}. The AP addresses this challenge through automated collection and evaluation of trainee actions, streamlined feedback for instructors/trainers (\textbf{Service Managers}), and a standard PE framework for coordinating data collection across multiple CRs~\cite{vsvabensky2024}.

ACTING relies on two components for automated performance analysis. The PE component, deployed at each participating CR, collects scenario data and maps trainee actions to trainee identities. The FPE component, operating at the AP level, aggregates the data received from all PE components, provides a unified view of Service Consumer performance, generates reports on scores and incorrect responses, and recommends additional units or services for further skill development.

The ACTING Federated PE framework is built around the concept of a \textit{Metric}, which represents a measurable criterion for assessing trainee actions. In CR scenarios, each Unit is associated with multiple metrics, defined during service design and adjustable during pre-execution by the Service Manager. Performance scores are computed through hierarchical aggregation: metrics are first combined into Unit-level assessments, which are then aggregated into a Service-level evaluation to produce the final PE score for each trainee or team.

\begin{figure}[htbp]
 \centering
 \includegraphics[width=\linewidth]{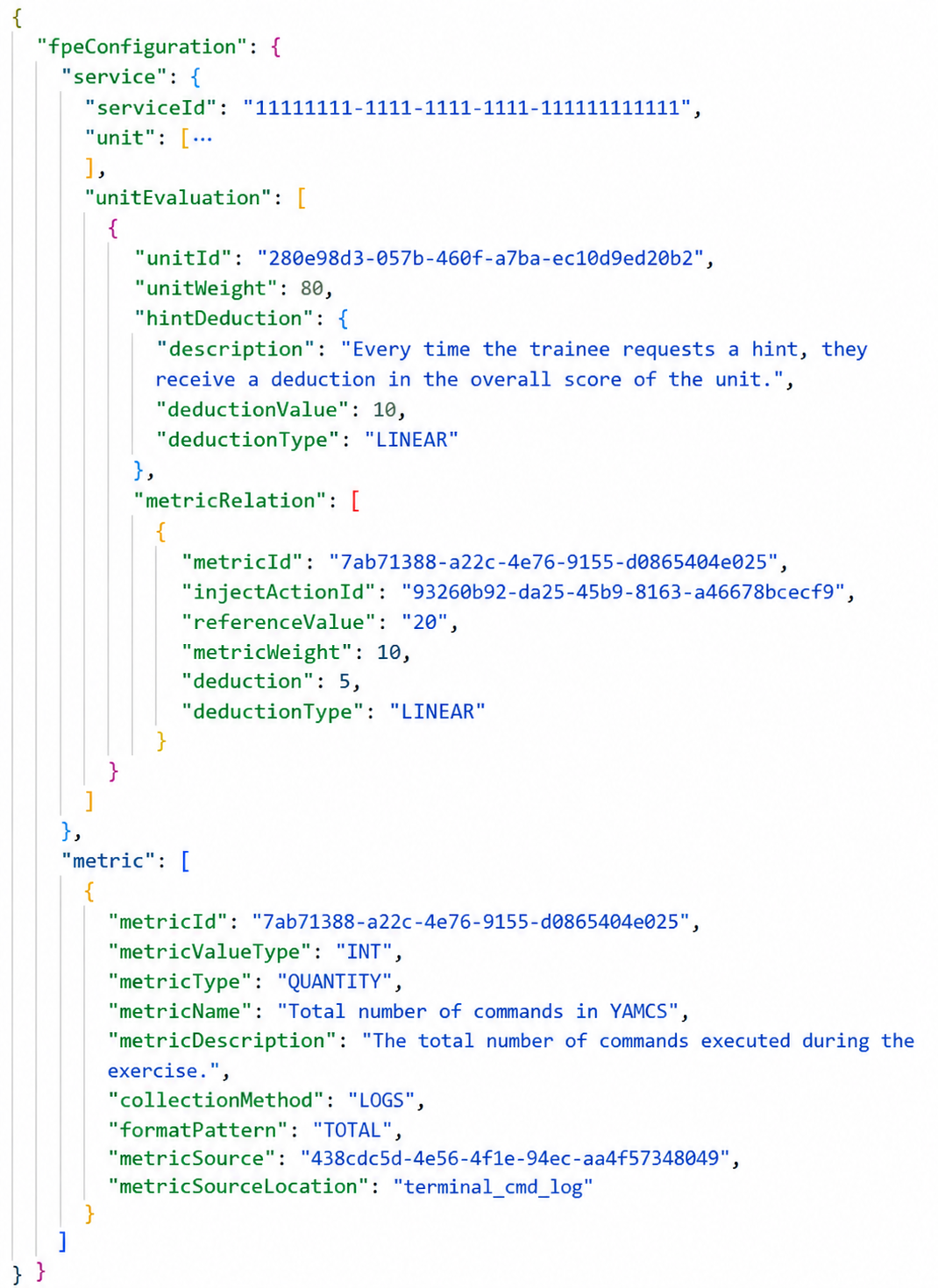}
 \caption{Snippet of service configurations required for FPE.}
 \label{fig:FPE_configs}
\end{figure}

For each service, the FPE component receives an EDL-FG configuration file describing the metrics to be evaluated, their reference solutions, and the scenario locations from which scoring data must be collected. Figure~\ref{fig:FPE_configs} shows a snippet of this configuration file.

Depending on the training objectives, FPE supports four metric types: (i) \textit{time}, which evaluates the duration of an activity, e.g., the time required to enable a firewall; (ii) \textit{quantity}, which evaluates the number of correctly executed actions, e.g., the number of commands; (iii) \textit{sequence}, which evaluates command execution and ordering against a predefined solution using fuzzy matching; and (iv) \textit{task}, which evaluates task completion, e.g., blocking a specific port.

For quiz-based exercises, the FPE component interacts with the LMS to retrieve completed quizzes, which are treated as another type of metric. Multiple-choice and true/false questions are graded by the LMS, while essay-type responses are evaluated by FPE using Natural Language Processing across three dimensions: (i) \textit{similarity}, which measures the Similarity Score (SimS) between the Service Consumer's answer and the reference solution, from 0, indicating low similarity, to 1, indicating identical responses; by default, FPE uses Sentence-BERT embeddings and cosine similarity; (ii) \textit{spelling}, which measures Spelling Score (SpellS) correctness, from 0, indicating that every word is incorrect, to 1, indicating no spelling mistakes; and (iii) \textit{clarity score (ClarS)}, which combines text assessment metrics, including Flesch Reading Ease (FRE), Connectivity (Conn), Subordinate Ratio (SubR), Average Maximal Depth (AvgMaxD), Mean Dependency Distance (MDD), LSA Cohesion (LSACoh), and CoLA-based quality (QCoLA), with Gaussian distributions used to account for non-linear behaviour. Specifically, FRE is a readability test that scores English text from 0 to 100, with higher scores indicating easier-to-read material; LSACoh estimates semantic coherence using Latent Semantic Analysis; and QCoLA estimates linguistic acceptability based on models trained on the Corpus of Linguistic Acceptability. ClarS is computed as follows:

\begin{equation*}
\begin{aligned}
\mathrm{ClarS}=\mathrm{QCoLA}^2\cdot\big(&w_1\mathrm{Conn}+w_2\mathrm{SubR}+w_3\mathrm{AvgMaxD}\\
&+w_4\mathrm{MDD}+w_5\mathrm{LSACoh}+w_6\mathrm{FRE}\big)
\end{aligned}
\end{equation*}

The final Essay Score (ES) is computed as:

\begin{equation*}
\mathrm{ES}=0.8\,\mathrm{SimS}+0.1\,\mathrm{SpellS}+0.1\,\mathrm{ClarS}
\end{equation*}

The weights were determined through experiments on the ASAG dataset\footnote{\url{https://github.com/DigiKlausur/ASAG-Dataset}}.

The resulting scores are displayed in near real time through the AP UI. Figure~\ref{fig:FPE_scores} shows the metrics of Scenario~2 and the respective scores of a trainee.

\subsection{ACTING Exercise Description Language - First Generation}
\label{ssec:apis}

\begin{figure*}[t]    
    \centering
    \begin{minipage}[t]{0.48\textwidth}
        \centering
        \includegraphics[scale=0.45]{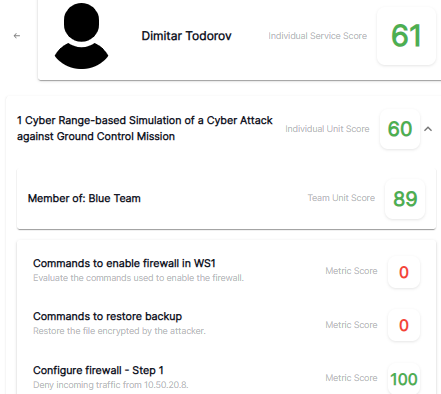}
        \caption{Visualisation of different metric scores for a trainee completing a unit of Scenario 2.}
        \label{fig:FPE_scores}

    \end{minipage}
    \hfill
    \begin{minipage}[t]{0.48\textwidth}
        \centering
        \includegraphics[width=\linewidth]{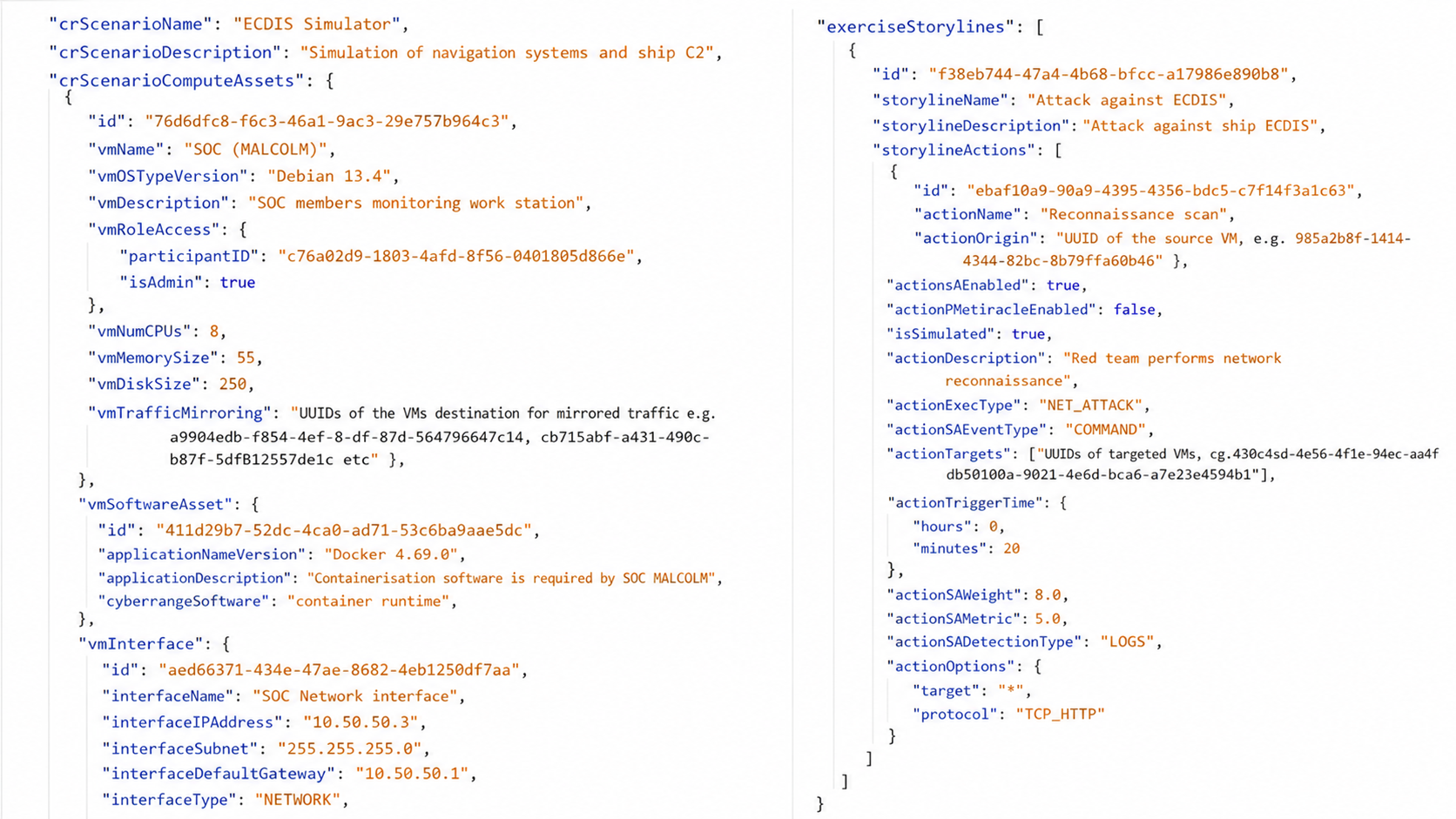}
        \caption{Sample of EDL-FG CR Scenario Description}
        \label{fig:topology_Storyline}
    \end{minipage}
\end{figure*}

The EDL-FG is a structured language for specifying cyber-range training services and exercises. It formally describes both the technical infrastructure required for simulation and the scenario logic governing cyber events and participant interactions. Specifically, EDL-FG captures two key dimensions: (i) computational and network resources used to simulate or emulate ICT/OT infrastructures and (ii) the scenario screenplay, including storylines, events, injects, and expected participant actions. By formalizing these elements, EDL-FG supports collaboration among cyber-range providers, trainers, and service consumers, while enabling automation in deployment, cyber-incident orchestration, and participant performance assessment~\cite{Yamin2022,Costa2020}. Figure~\ref{fig:topology_Storyline} presents an excerpt from an EDL-FG file generated from a real scenario, illustrating representative assets from the topology and the corresponding event--inject--action chain.

EDL-FG is implemented as an extensible YAML-based data structure, selected for its readability, flexibility, and compatibility with modern orchestration frameworks. The language supports hierarchical object definitions and extensible schemas, enabling the gradual inclusion of new attributes and entities as cyber-range capabilities evolve. Through YAML schemas aligned with JSON Schema standards, EDL-FG enables automated validation of exercise descriptions, ensuring structural consistency and correct data typing. The data model represents multiple abstraction layers, including services, training units, infrastructure assets (e.g., compute, network, and custom assets), and scenario elements such as storylines, events, and injects. This structure allows complex and cascading cyber simulations to be defined consistently across single or federated cyber-range environments.

The ACTING EDL-FG is deliberately designed as a unifying semantic layer that bridges multiple established cybersecurity frameworks and standards. Rather than reinventing concepts, it reuses and refers to classes, taxonomies, and labels from these frameworks and extends them with additional semantics, cross-references, and domain-specific literals tailored to the AP specifics. Such concepts include:
\begin{itemize}
    \item \textit{Cyber Data Exchange Model (CDEM)} and \textit{BONES}~\cite{DEMBONES}: EDL-FG adopts foundational object classes and relationships for representing cyber environments, such as assets, networks, communication flows, and observable effects. 
    \item \textit{MITRE ATT\&CK}~\cite{MITREATTACK} and \textit{MITRE DEFEND}~\cite{MITREDEFEND} contribute standardised vocabularies for adversarial behaviours and defensive countermeasures. Within EDL-FG, these vocabularies are used to formally annotate scenario actions, attack paths, and mitigation strategies, enabling precise mapping between simulated events and real-world tactics, techniques, and procedures.
    \item \textit{Common Weakness Enumerations (CWEs)}~\cite{CWE} is used to describe vulnerabilities as structured weaknesses, allowing the scenario designer to link system misconfigurations or software flaws to prerequisites for executing the planned events. 
    \item \textit{High-Level Architecture}~\cite{HighLevelArchitecture} principles inform the modelling of distributed simulation components and interactions, particularly for synchronising simulation entities, e.g., red team tools, blue team monitoring systems, and physical devices, within a federated CR.
    \item \textit{TOSCA}~\cite{TOSCA} contributes concepts for service topology and orchestration, which EDL-FG leverages to describe how compute assets, services, and dependencies are deployed, configured, and managed across the CR infrastructure.    
    \item The \textit{NIST NICE framework}~\cite{NISTNICE} provides a structured approach for representing the relationships among roles, skills, and learning objectives.
\end{itemize}

What distinguishes ACTING EDL-FG is not just the aggregation of these frameworks, but the way it interconnects them through explicit cross-references and validation rules. For example, a single scenario element---such as a phishing attack implemented on a specific enterprise architecture---can simultaneously reference: (i) \textit{mitre attack technique}; (ii) \textit{associated cwes representing exploited weaknesses}; (iii) \textit{defensive measures from mitre defend}; (iv) \textit{affected assets defined via CDEM}; and (v) \textit{roles responsible for detection and response via nice}.

In addition, EDL-FG introduces new literals and extensions to capture CR service--specific requirements that are not fully addressed by existing standards. These include: (i) \textit{temporal and conditional constructs for orchestrating events (triggers)}; (ii) \textit{interaction options between simulated users and simulated systems}; (iii) \textit{evaluation metrics and sa scoring mechanisms}; and (iv) \textit{abstractions for hybrid (simulated and emulated) IT/OT components}.

By combining standardised vocabularies with these extensions, EDL-FG creates a shared, formalised language that improves communication among multidisciplinary stakeholders, such as cybersecurity experts, scenario designers, educators, and system engineers. It ensures that scenario design, execution, and evaluation are consistent, reproducible, and semantically aligned with real-world cybersecurity knowledge and practices. In combination with the SDL-T, EDL-FG establishes a structured framework for the design, validation, and execution of cybersecurity training services. By separating the conceptual description of exercises from the execution logic, this approach enables automation, interoperability across federated CRs, and improved reproducibility of cybersecurity training scenarios, while supporting the future evolution of EDL-FG towards a standardised language for describing cyber exercises and simulation-based cybersecurity training environments.

\subsection{User Simulator}
\label{ssec:apis}
The User Simulator (US) resides within the CR and simulates benign and malicious human-generated network traffic. Simulated profiles may represent: (a) Blue users, who exhibit benign behaviour and respond to social-engineering and network events based on predefined profiles, such as basic or advanced IT users; (b) Red users, who act as malicious actors performing social-engineering or network-based attacks, such as phishing or distributed denial-of-service (DDoS) attacks; and (c) Grey users, who perform neutral activities such as file creation or web browsing.

Before exercise execution, the simulated-user configuration is defined, including the number of users, their roles, and the actions each user will perform with corresponding timestamps. During execution, users follow the scenario timeline, and after each action, relevant information is forwarded to the FA component for reporting to the trainer. Through the FAE user interface, trainers may also dynamically modify simulated-user timelines or behaviours during the exercise.

\section{Cross-Domain Training Scenarios}
\label{sec:scen}

This section presents three representative cybersecurity training scenarios (sub-sections \ref{ssec:sc1}, \ref{ssec:sc2}, \ref{sec:sc3}) implemented within the AP. These scenarios illustrate how stakeholders can design, orchestrate, and execute federated cyber-range exercises within the ACTING ecosystem. The section demonstrates the platform’s operational capabilities and flexibility in supporting diverse cyber-defence training environments, focusing on the types of exercises that can be deployed rather than the detailed scenario-development methodology. It also highlights the platform’s ability to support complex, multi-partner training activities across the exercise lifecycle, from scenario conception to coordinated execution.

\subsection{Scenario 1: Combined cyber-attacks against joint HQ, Land and Navy CIS systems}
\label{ssec:sc1}

In Scenario 1, the network infiltration phase uses social engineering through strategically placed USB devices containing malware designed to exploit autorun vulnerabilities. Using PowerShell-based intrusion techniques, the malware establishes covert access to compromised hosts and enables lateral movement across the Command and Control (C2) network, allowing the adversary to manipulate operational data across interconnected systems. After persistence is established, the attacker modifies shared calendars, tampers with logistics databases, disrupts maintenance schedules, and falsifies asset coordinates in the Common Relevant Picture, resulting in degraded SA, disrupted unit coordination, and reduced mission readiness.

\begin{figure}[htbp]
\centering
\includegraphics[width=\linewidth]{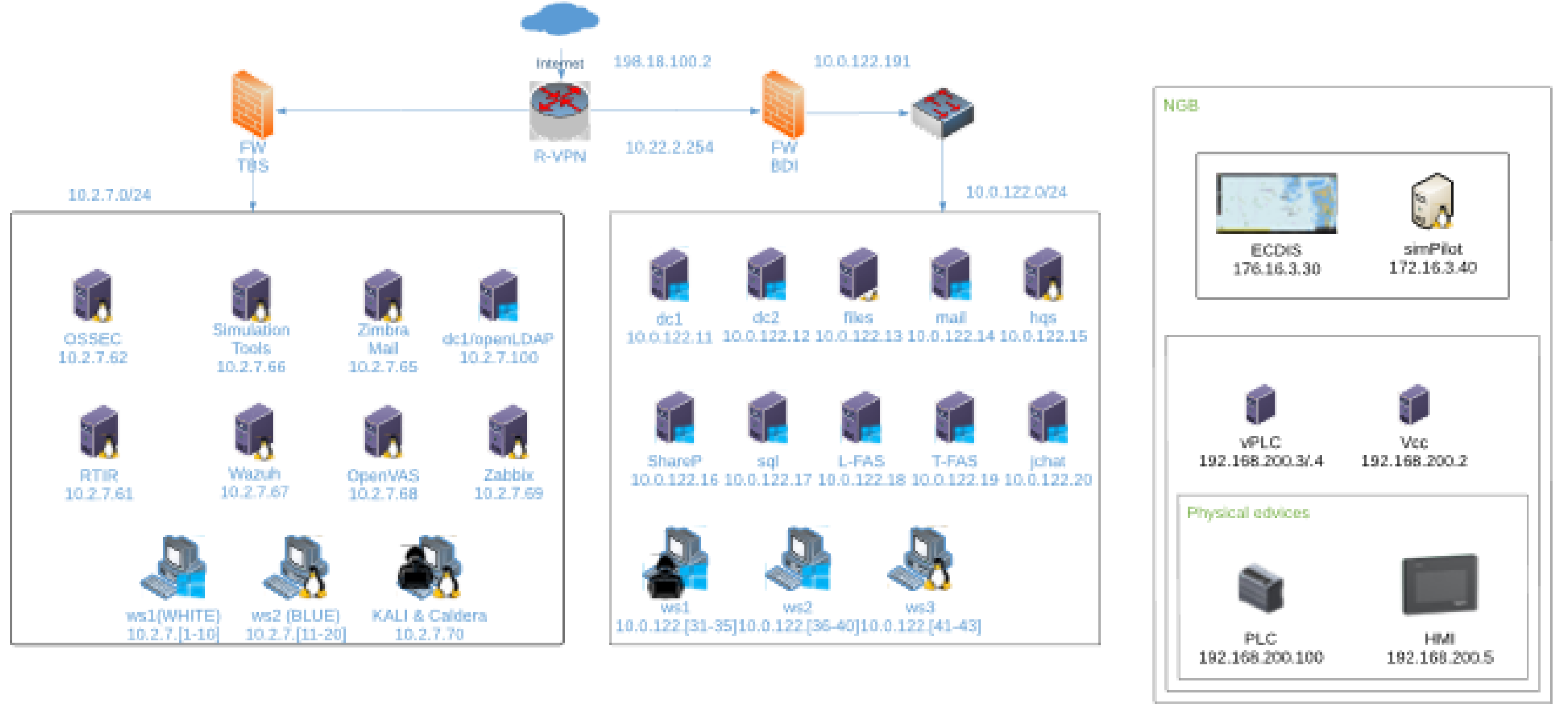}
\caption{ACT-SC-01 network topology diagram}
\label{fig:ACT-SC-01}
\end{figure}

\subsection{Scenario 2: Space and Maritime Sectors}
\label{ssec:sc2}

The ACT-SC-02 scenario presents a refined cyber-physical attack that demonstrates the complex interplay between space-based infrastructure and maritime operations. Through its two interconnected storylines—GNSS signal manipulation and a subsequent ransomware attack—the scenario highlights how modern threat actors can exploit vulnerabilities across multiple sectors to produce cascading operational effects. By mapping these events to established cybersecurity frameworks, the scenario provides a comprehensive training environment that enables satellite operators and maritime personnel to detect, respond to, and mitigate such advanced threats. This realistic approach to CD training enhances the resilience of critical infrastructure while promoting cross-sector collaboration and strengthening incident response capabilities.

\begin{figure}[htbp]
\centering
\includegraphics[scale=0.13]{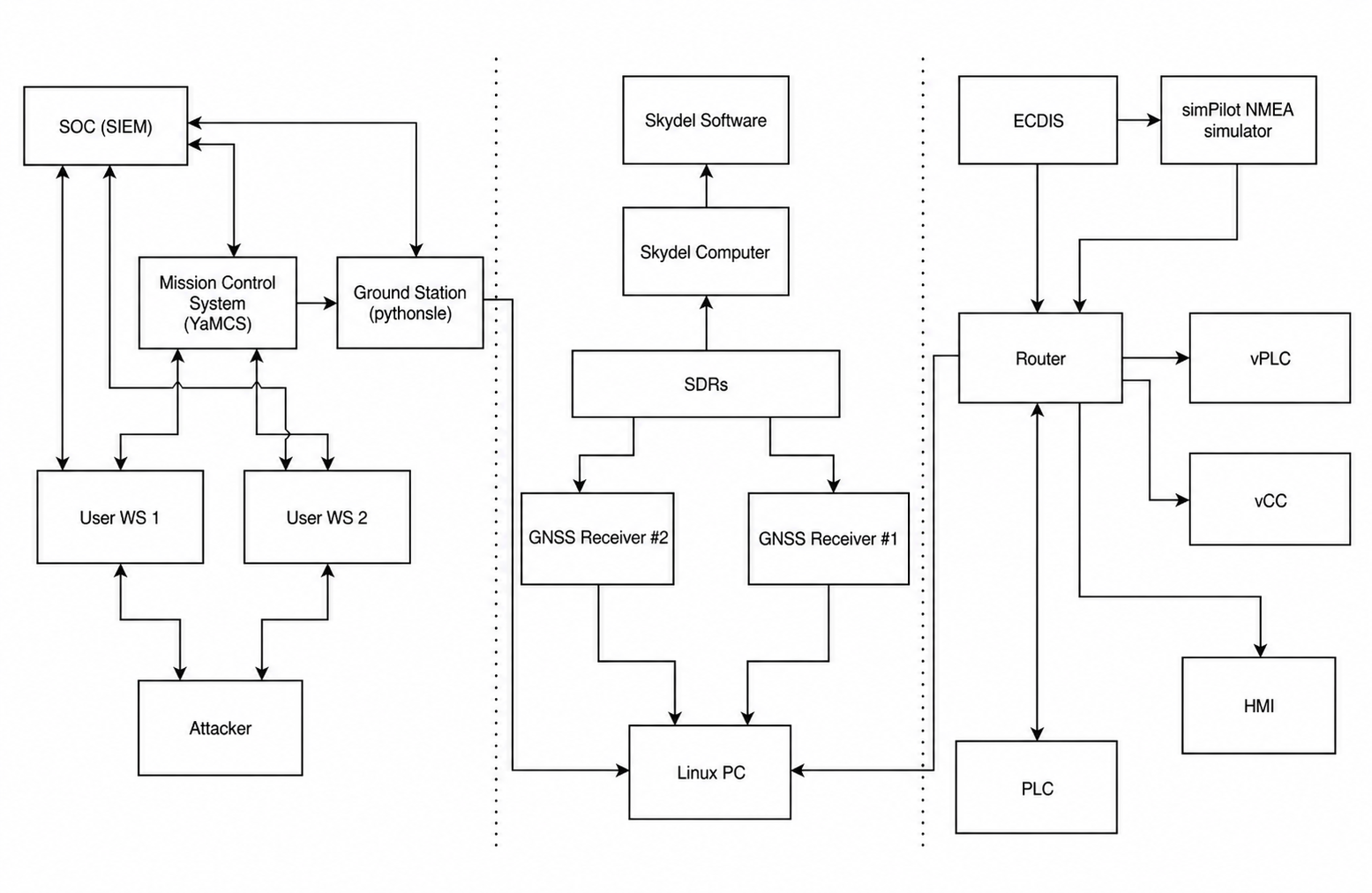}
\caption{ACT-SC-02: Cyber-range environment}
\label{fig:ACT-SC-02}
\end{figure}

\subsection{Scenario 3: Military Base Classified Information Theft}\label{sec:sc3}

This fictional training scenario focuses on detecting and containing a simulated targeted breach in a high-security military environment. Following an initial foothold, the exercise focuses on identifying anomalous authentication patterns and lateral movement across a simulated internal network. Participants should use representative Military Command \& Control (C2) servers and SIEM aggregation tools to analyse deviations in access-control logs and audit trails. The technical challenge lies in distinguishing between standard operational traffic and unauthorised access to fictional classified data repositories, especially as white-team injects escalate the severity of the simulated data exposure.

The primary objectives involve identifying potentially compromised accounts and tracing the simulated attacker’s path through secure access segments. Once the breach scope is established, participants should isolate affected assets while following representative military Standard Operating Procedures (SOPs). This includes technical mitigation and formal escalation through a simulated chain of command, ensuring that grid security and military intelligence stakeholders are informed about the integrity of the fictional classified systems.

\begin{figure}[htbp]
\centering
 \includegraphics[width=\linewidth]{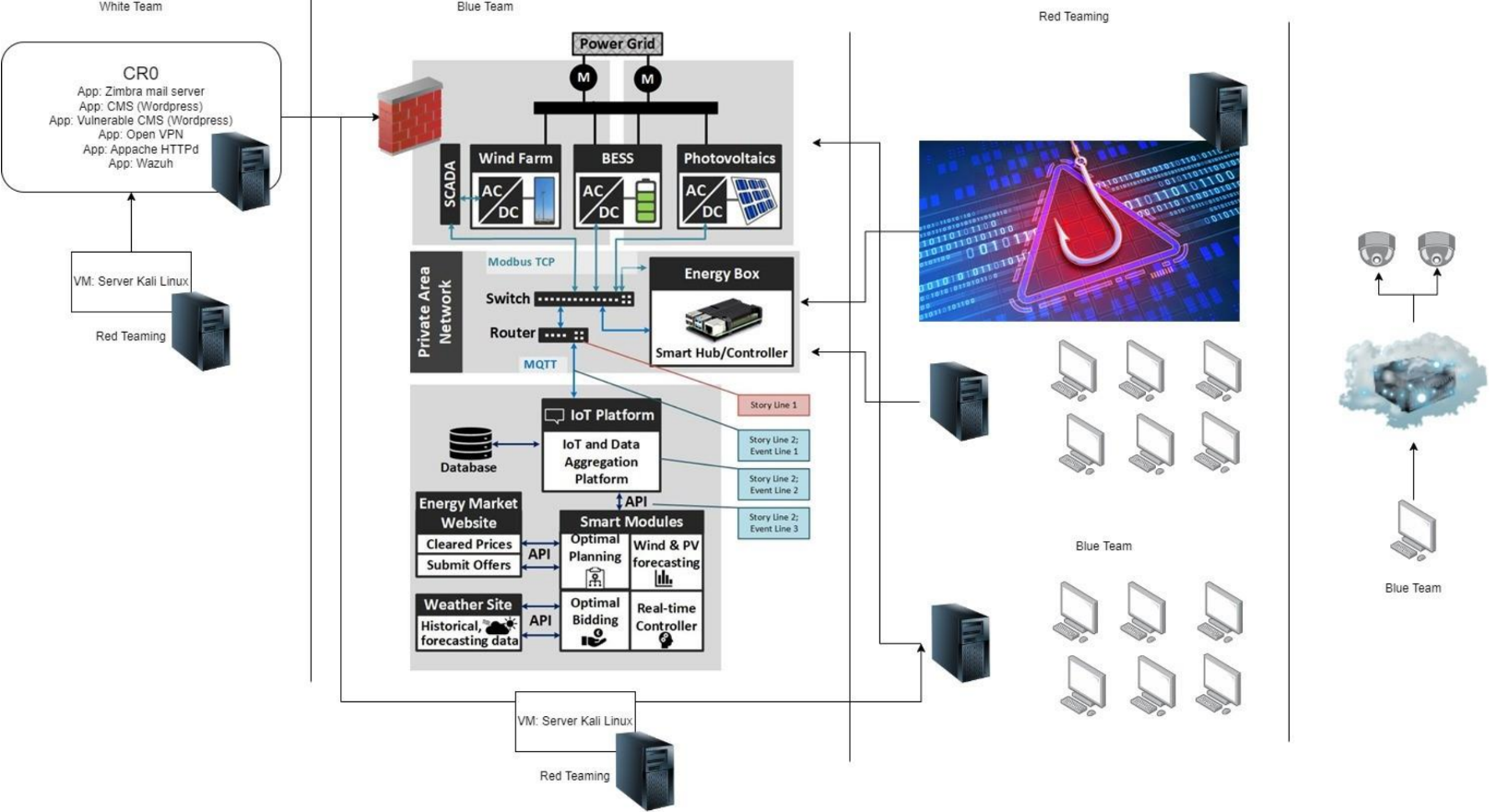}
\caption{ACT-SC-03: Cascading effect for cyber-attack in the civilian sector}
\label{fig:ACT-SC-03}
\end{figure}

\section{Conclusions}
\label{sec:conc}

This paper presented three key contributions to interoperability and scalability in cyber-range training environments. First, it introduced EDL-FG, a structured approach for describing cyber-range exercises through infrastructure configuration and scenario logic. Second, it presented the AP’s automated PE and scoring mechanisms for assessing trainee actions across participating CRs. Third, it demonstrated support for multi-domain cyber-training scenarios combining civilian and military contexts. Overall, ACTING advances federated cyber training through automated PE, SA, and coordinated scenario execution across distributed infrastructures. Future work will include systematic empirical evaluation with quantitative measurements, baselines, and validation, while further developing EDL-FG into a mature exercise-description language and introducing adaptive scenarios that adjust complexity based on trainee performance.

\bibliographystyle{IEEEtran}
\bibliography{csr2026-acting}

\end{document}